\documentstyle[subeqn]{article}
\begin{document}
\baselineskip = 18 pt

\begin{titlepage}
\begin{center}
\large {A NEW CLASS OF OPTICAL SOLITONS}
\end{center}
\vspace{4 cm}
\begin{center}
{\bf Sasanka Ghosh}\\
{\it Physics Department, Indian Institute of Technology, Guwahati}\\
{\it Panbazar, Guwahati - 781 001, India}\\
{\it Phone: +91-361-521-915, Fax: +91-361-521-916}\\ 
{\it E-mail: sasanka@iitg.ernet.in}\\
\end{center}
\begin{center}
and
\end{center}
\begin{center}
{\bf Sudipta Nandy}\\
{\it Physics Department, Indian Institute of Technology, Guwahati}\\
{\it Panbazar, Guwahati - 781 001, India}\\
{\it Phone: +91-361-521-915, Fax: +91-361-521-916}\\ 
{\it E-mail: sudipta@iitg.ernet.in}\\
\end{center}

\end{titlepage}
\newpage
\baselineskip = 18 pt

{\bf ABSTRACT}

Existence of a new class of soliton solutions is shown for higher order 
nonlinear Schr{\" o}dinger equation, describing third order dispersion, Kerr 
effect and stimulated Raman scattering. These new solutions have been obtained 
by invoking a group of nonlinear transformations acting on localised stable 
solutions. Stability of the solutions has been studied for different values of 
the arbitrary coefficient, involved in the recursion relation and consequently, 
different values of coefficient lead to different transmission rate for almost 
same input power. Another series solution containing even powers of localised 
stable solution is shown to exist for higher order nonlinear schr{\" o}dinger 
equation. 

\vspace{1 cm}


\vspace {2 cm}

\baselineskip = 24 pt

Theoretical prediction of Hasewaga and Tappert that an optical pulse in a 
dielectric fibre forms an envelop soliton \cite {1} and subsequent experimental 
verification by Mollenauer \cite{2} have made a significant acheivement in 
ultra-high
speed telecommunication. Contrarary to conventional optical pules, solitonic 
optical pulses are dispersion less, which makes them so important for long 
distance communication. Optical solitons are currently under intense study not 
only for providing distortsion less signal transmission, but also for many 
nonlinear optical applications in fibre.

Nonlinear Schr{\" o}dinger equation (NLS) and their various higher order 
generalizations play a pivotal role in describing the dynamics of optical 
solitons. The successful propagation of NLS solitons for long distance 
transmission eventually turns out to be limited to a few kilometers \cite {2}. 
This is due to the fact that 
solitons, which follow the dynamics of NLS cannot take care of propagation loss, 
as it propagates along the optical fibre and consequently the pulse needs to be 
reshaped time to time to compensate propagation loss. As a remedy, at present, a 
Raman pump light is used to send a signal simultaneously along the fibre to 
adjust the propagation loss \cite {3,4}. This is acheived by utilizing the 
effect of stimulated Raman scattering. It is seen later \cite {5,5a,6} that the 
effect of stimulated Raman 
scattering already exists withing the spectra of soliton solutions, if we 
include a few more terms in the equation of motion over NLS. No additional pump 
light is needed in the later case to adjust the propagation loss. The equation 
of motion, thus obtained, is called higher order nonlinear Schr{\" o}dinger 
equation (HNLS):
\begin{equation}
\partial_z E = i[\alpha_1 \partial_{\tau \tau} E + \alpha_2 \vert E \vert ^2 E] 
+ \epsilon [\alpha_3 \partial_{\tau \tau \tau} E 
+ \alpha_4 \partial_{\tau} (\vert E \vert ^2 E) 
+ \alpha_5 \partial_{\tau} (\vert E \vert ^2) E]
\label{I1}
\end{equation}
where, $E$ is envelop of electric field propagating in $z$ direction at a time
$\tau$. Notice that in absence of last three terms, (\ref {I1}) reduces to NLS. 
The coefficients of the higher order terms, namely $\alpha_3$, $\alpha_4$ and
$\alpha_5$ respectively  represent third order dispersion (TOD), self-steeping
(SS) related to Kerr effect and the self frequency shifting via stimulated Raman 
scattering. It is the last term, which plays an important role in the 
propagation of distortsion less optical pulses over long distance. On the 
other hand, for ultra-short pulses, the effects of last three terms cannot be 
neglected. 

Several attempts are made to find the soliton solutions of HNLS (\ref {I1}) 
\cite [8-12]{} by various methods like Hirota's bilinear method, Painleve analysis, Backlund 
transformation, inverse scattering technique. Recently, HNLS is solved for 
soliton solutions by inverse scattering technique method for an arbitrary value 
of the coefficients \cite {12}, where it is convenient to solve a gauge equivalent 
equation,
\begin{equation}
\partial_{t}q(x,t) = \epsilon[\partial_{xxx}q(x,t) + \gamma_1 \vert q(x,t)
\vert^2 \partial_x q(x,t) + \gamma_2 \partial_x(\vert q(x,t)\vert^2)q(x,t)]. 
\label{I2}
\end{equation}
The equations (\ref {I1}) and (\ref {I2}) are related explicitely through the 
following gauge transformation: 
\begin{subequations}
\begin{eqnarray}
E(z,\tau) &=& exp[ -ipx + i\epsilon p^3t ] q(x,t)\\
\label{IIIa}
x &=& \tau + \alpha_1pz\\
\label{IIIb}
t &=& \alpha_3z
\label{IIIc}
\end{eqnarray}
\label{III}
\end{subequations}
with $p=\alpha_1/(3\epsilon \alpha_3)=\alpha_2/(\epsilon \alpha_4)$, $\gamma_1=
\alpha_4/\alpha_3$, and $\gamma_2=\alpha_5/\alpha_3$.  
 
One soliton solution, found by inverse scattering method, may be given by 
\cite{12} 
\begin{equation}
E(z,\tau) = (\eta/{\sqrt(n-1)})sech {\tilde A}(z,\tau) exp(i{\tilde B}(z,\tau))
\label{XIII}
\end{equation}
with 
\begin{eqnarray*}
{\tilde A}(z,\tau) &=& \eta \tau + [\epsilon \alpha_3(\eta^3 - 3\xi^2 \eta )
 + {\alpha_1^2\over{3\epsilon \alpha_3}}\eta ]z - \gamma - {1\over 2} ln(n-1)\\
{\tilde B}(z,\tau) &=& [\xi - {\alpha_1\over{3\epsilon \alpha_3}}] \tau 
+ [-\epsilon \alpha_3(\xi^2 - 3 \eta^2 
- {\alpha_1^2\over {3\epsilon^2\alpha_3^2}})\xi 
- {2\alpha_1^3\over{27\epsilon^2 \alpha_3^2}}]z + \delta 
\end{eqnarray*}
where, $n$ is dimension of the Lax operator and the location of the simple 
pole is chosen at $\lambda=\frac{1}{2}(-\xi+i\eta)$ in the complex spectral 
parameter plane, $\xi$ and $\eta$ being real parameters. Simple {\it sech} type 
structure of one soliton in (\ref {XIII}) is quite interesting, since it  may be 
obtained from the output of a mode locked laser. $N$ soliton solutions is 
obtained for a more a general dynamical equation in \cite {13}. 

In this paper, we discuss a new class of solution of (\ref {I1}). For this, let 
us assume a trial solution of (\ref {I1}) in the form
\begin{equation}
E(z,\tau)= E_0e^{ilz+im\epsilon}y(\tau)
\label{I5}
\end{equation}
where $l$ and $m$ are real parameters and $y(\tau)$ is chosen as a real 
function. Substituting (\ref {I5}) in (\ref {I1}), we obtain a set of two
differential equations in $y$: 
\begin{subequations}
\begin{equation}
(-l-\alpha_1m+\epsilon\alpha_3m^3)y(\tau) + (\alpha_1-3\epsilon\alpha_3m)
y^{\prime \prime}(\tau)+(\alpha_2-\epsilon\alpha_4m)E_0y^3(\tau)=0
\label{I6a}
\end{equation}
\begin{equation}
(2\alpha_1m-3\epsilon\alpha_3m^2)y^{\prime}(\tau)
+\epsilon\alpha_3y^{\prime \prime \prime}(\tau)+\frac{\epsilon}{3}(\alpha_4
+2\alpha_5)(y^3)^{\prime}(\tau)=0
\label{I6b}
\end{equation}
\label{I6}
\end{subequations}
Consistency conditions of (\ref {I6a}) and (\ref {I6b}) demands that
\begin{eqnarray}
\alpha_1&=&\epsilon(1+3m)\alpha_3\\
l&=&-2\epsilon m(1+m)^2\alpha_3\\
\alpha_4+2\alpha_5&=&\frac{3E_0^2}{\epsilon}[\alpha_2-\epsilon m\alpha_4]
\label{I7}
\end{eqnarray}
Notice that (\ref {I6}) leads a solution of the form
\begin{equation}
y=sech(E_0\tau)
\label{I8}
\end{equation}
provided the following conditions are satisfied
\begin{eqnarray}
m&=&\frac{\alpha_1-\epsilon\alpha_3}{3\epsilon\alpha_3}\\ \nonumber
l&=&-\frac{2(\alpha_1-\epsilon\alpha_3)(\alpha_1+2\epsilon\alpha_3)^2}
{\epsilon^2\alpha_3^2}\\ \nonumber
E_0^2&=&-\frac{(\alpha_1-3\alpha_3)(\alpha_1+\epsilon\alpha_3)}
{3\epsilon^2\alpha_3^2}\\ \nonumber
\alpha_2&=&2\epsilon\alpha_3-\epsilon\alpha_4+\frac{\alpha_1\alpha_4}
{3\alpha_3}\\ \nonumber
\alpha_5&=&\frac{(3\alpha_3-\alpha_1)(\alpha_1+\epsilon\alpha_3)}
{\epsilon^2\alpha_3}-\frac{\alpha_4}{2}
\label{I8a}
\end{eqnarray}
{\it sech} type solution of (\ref {I6}) is a desired one, as it has been 
obtained earlier \cite {12} by inverse scattering method (\ref {XIII}). In this 
paper, however, we look for a more general solution of the form
\begin{equation}
y=\sum_{j=0}^{\infty} C_j y^j(\tau)
\label{I9}
\end{equation}
Substituting (\ref {I9}) into (\ref {I6}), we obtain
\[ 
C_0=0,\quad\quad \pm \sqrt (\frac{l+m^2\alpha_1-\epsilon m^3\alpha_3}
{(\alpha_2-\epsilon m\alpha_1^2)E_0^2})  
\]
Unfortunately, nontrivial values of $C_0$ lead only to trivial solutions of
(\ref {I6}) consistent with the conditions (\ref {I8a}). Thus $C_0=0$ is an 
acceptible solution and consequently the 
recursion relation among the coefficients $C_j$ may be given by
\begin{eqnarray}
&~&[(-l-m^2\alpha_1+\epsilon m^3\alpha_3)+q_0^2(\alpha_1-3\epsilon m\alpha_3)
(j+2)^2]C_{j+2}\\ \nonumber
&=&q_0^2(\alpha_1-3\epsilon m\alpha_3)(j^2+j)C_j-q_0^2(\alpha_2-
\epsilon m\alpha_4)\sum_{m=0}^{j+2}\sum_{l=0}^{m}C_{j+2-m}C_{m-l}C_{l} \\ 
\nonumber
&~&
\label{I9a}
\end{eqnarray} 
for $j=0,1,2,3,\cdots$. If we now choose $C_1$ arbitrary, all the even 
coefficients become trivial
\begin{equation}
C_{2j}=0
\label{I10}
\end{equation}
whereas odd coefficients satisfy a recursion relation 
\begin{equation}
C_{2j+1}=\frac{1}{2j(j+1)}[j(2j-1)C_{2j-1}-\sum_{m=0}^{2j-1}\sum_{l=0}^{m}
C_{2j+1-m}C_{m-l}C_{l}
\label{I11}
\end{equation}
provided the coefficients satisfy the conditions (\ref {I8a}), which is not
surprising, as we will see $y=sech(E_0\tau)$ itself is a special solution of 
(\ref {I9}). It is interesting to observe from the recursion relation 
(\ref {I11}) that if the arbitrary coefficient $C_1$ is chosen to be unity all
higher order coefficients starting from $C_3$ become zero and the series leaves
with one term, $y=sech(E_0\tau)$. Thus the solution (\ref {I8}) already exists
as a special case in (\ref {I9}) for nontrivial odd coefficients. Notice that a
similar solution also exists for NLS \cite {14}. 

Let us we now introduce an operatorial notation for the series solution 
(\ref {I9}) as
\begin{equation}
{\bf H}_{C_1}y=\sum_{j=1}^{\infty}C_{2j+1} y^{2j+1}
\label{I12}
\end{equation}
for arbitrary $C_1$ such that  
\[ {\bf H}_1y=y\]
{\it i.e.} the coefficients satisfy the recursion relations (\ref {I10}) and 
(\ref {I11}). It emerges from (\ref {I12}) and the recursion relations that 
\begin{equation}
{\bf H}_{C_1}{\bf H}_{C'_1}={\bf H}_{C_1C'_1}
\label{I13}
\end{equation}
implying that $\{{\bf H}_{C_1};C_1\neq0\}$ is the commutative multiplicative group of 
nonlinear transformation (GNT). Thus, like NLS, the commutative group of 
nonlinear transformations for HNLS also leads to a new solution by operating on 
stationary localised solutions. The advantage of GNT method is that it enables 
us to determine the stability of the new solution by the following relation 
\begin{equation}
lim_{\epsilon \rightarrow 0} {\bf H}_{1+\epsilon}y=y
\label{I14}
\end{equation} 
The main motivation in obtaining new type of solutions by GNT method is to 
increase the transmission rate. It is observed in NLS that higher values of 
the coefficient $C_1$ gives rise to higher transmission rate for almost the same
input power \cite {14}. Interestingly, the new form of solutions (\ref {I10}) 
and (\ref {I11}) also assert the same conclusion for HNLS with an added 
advanatge that HNLS takes care of propagation loss, unlike NLS. 

Let us present another interesting solution for HNLS, which emerges from the 
recursion relation (\ref {I9a}). If we choose $C_2$ arbitrary and impose 
the condition that
\begin{equation}
2(-l-m^2\alpha_1+\epsilon m^3\alpha_3)=q_0^2(\alpha_1-3\epsilon m\alpha_3)
\label{I15}
\end{equation}
all odd coefficients will become trivial, {\it i.e.}
\begin{subequations}
\begin{equation}
C_{2j+1}=0
\label{I16a}
\end{equation}
whereas even coefficients will satisfy a recursion relation, given by 
\begin{equation}
C_{2j}=\frac{2}{(8j^2+1)}[2(j-1)(2j-1)C_{2j-2}
-\frac{\alpha_2-\epsilon m\alpha_4}{\alpha_1-3\epsilon m\alpha_3}
\sum_{m=0}^{2j}\sum_{l=0}^{m}C_{2j-m}C_{m-l}C_l]
\label{I16b}
\end{equation}
\label{I16}
\end{subequations}

To conclude, a new class of soliton solutions are obtained for HNLS. This is 
found by invoking a group of nonlinear transformations acting on a stable 
localised solution. It turns out that the new solution reduces to the most 
stable one for $C_1=1$. For higher values of $C_1$, however, it leads to a 
faster rate 
of propagation for an almost same input power. This ,indeed, is a considerable 
improvement in the field of all optical communication through fibre.  The 
existence of another solution of HNLS, where all odd coefficients are trivial, 
also earns a lot of interest and needs to be investigated more thoroughly. It 
is quite useful to see the consequences of the even series solution in the 
context of fibre communication. 

\vspace{1 cm}

{\it S.N. would like to thank CSIR, Govt. of India for fellowship and financial
support.}


\vspace {2 cm}

\baselineskip 18 pt
 
\begin{thebibliography}{99}
\bibitem{1} A. Hasewaga and F.D. Tappert, Appl. Phys. Lett. 23, 142 (1973).
\bibitem{2} L.F. Mollenauer, R.H. Stolen and J.P. Gordon, Phys. Rev. Lett 45, 
1095 (1980).
\bibitem{3} A. Hasegawa, Appl. Opt. 23, 3302 (1984); Optical Solitons in Fibres,
 (springer, Heidelberg, 1989).
\bibitem{4} L.F. Mollenauer, K. Smith, Opt Lett. 13, 675 (1988).
\bibitem{5} J.P. Gordon, Opt. Lett. 11, 662 (1986).
\bibitem{5a} F.M. Mitschke and L.F. Mollenauer, Opt. Lett. 11, 657 (1986). 
\bibitem{6} Y. Kodama and A. Hasegawa, IEEE J. Quantum Electron. QE-23,5610 
(1987). 
\bibitem{7} N. Sasa and J. Satsuma, J. Phys. Soc. Jpn. {\bf 60}, 409 (1991).
\bibitem{8} K. Porsezian and K. Nakkeeran, Phys. Rev. Lett. 76, 3955 (1996).
\bibitem{9} M. Gedalin, T.C. Scott and Y.B. Band, Phys. Rev. Lett. 78, 448 
(1997).
\bibitem{10} K. Porsezein, M. Daniel and M. Lakshmanan, Proc. of the Int. Conf.
on Nonlinear Evolution Equations and Dynamical Systems, Eds. V.G. Makhankov, 
I. Puzynin and O. Pashaev 436 (World Sc., Singapore).
\bibitem{11} S. Ghosh, A. Kundu and S. Nandy, Soliton Solutions, Liouville 
Integrability and Gauge Equivalence of Sasa Satsume Equation, J. Math. Phys. 40, 
1993 (1999).
\bibitem{12} S. Ghosh and S. Nandy, Optical Solitons in Higher Order Nonlinear 
Schr{\" o}dinger Equation, Solv-int 9904019.
\bibitem{13} S. Ghosh and S. Nandy, Inverse Scattering Method and Vector Higher
Order Nonlinear Schr{\" o}dinger Equation, Solv-int 9904021.
\bibitem{14} S. Lekic, S. Galamic and Z. Rajilic, Optical Fibre Communication: 
Group of Nonlinear Transformations, Solv-int 9812012.
\end{thebibliography}
\end{document}